\documentstyle[preprint,aps,pra]{revtex}
\tightenlines
\begin{document}

\baselineskip 22pt

\draft

\title
{Hydrodynamic damping in trapped Bose gases}

\author{T. Nikuni}
\address{Department of Physics, Tokyo Institute of Technology,
Oh-okayama, Tokyo, 152 Japan \\
and Department of Physics, University of Toronto, Toronto,
Ontario, Canada M5S 1A7} 

\author{A. Griffin}
\address{Department of Physics, University of Toronto, Toronto,
Ontario, Canada M5S 1A7} 

\date{\today}

\maketitle

\begin{abstract}
Griffin, Wu and Stringari have derived the hydrodynamic
equations of a trapped dilute Bose gas above the Bose-Einstein transition
temperature.
We give the extension which includes hydrodynamic
damping, following the classic work of Uehling and Uhlenbeck based on
the Chapman-Enskog procedure.
Our final result is a closed equation for the velocity fluctuations
$\delta {\bf v}$ which includes the hydrodynamic damping due to the shear
viscosity $\eta$ and the thermal conductivity $\kappa$.
Following Kavoulakis, Pethick and Smith, we introduce a spatial
cutoff in our linearized equations when the density is so low that the
hydrodynamic description breaks down.
Explicit expressions are given for $\eta$ and $\kappa$, which are
position-dependent through dependence on the local fugacity when one includes
the effect of quantum degeneracy of the trapped gas.
We also discuss a trapped
Bose-condensed gas, generalizing the work of Zaremba,
Griffin and Nikuni to include hydrodynamic damping
due to the (non-condensate) normal fluid.
\end{abstract}


\clearpage

\section{Introduction}\label{sect1}
Recent observation of Bose-Einstein condensation in trapped atomic
gases has stimulated interest in the collective oscillations
of non-uniform Bose gases \cite{mewes,jin97}.
The hydrodynamic equations of a trapped Bose gas (neglecting damping)
have been recently derived
above \cite{gri97} and below \cite{zgn} 
the Bose-Einstein transition temperature ($T_{\rm BEC}$).
These hydrodynamic equations describe low frequency phenomena
($\omega \ll 1/\tau_c$, where $\tau_c$ is the mean time between collisions
of atoms), when collisions are sufficiently strong to ensure local
thermodynamic equilibrium.
These results can be used to derive closed equations for the velocity
fluctuations in a trapped gas, whose solution gives the normal modes of
oscillation.
In particular, above $T_{\rm BEC}$, Ref.~\cite{gri97} gives explicit results
for the frequencies of surface, monopole, and coupled 
monopole-quadrupole modes.
Available experimental data on the frequencies and damping of collective
oscillations in trapped gases \cite{mewes,jin97} is mainly for the
collisionless regime, rather than the hydrodynamic regime we consider in
this paper.
However sufficiently high densities which allow one to probe the
hydrodynamic region at finite temperatures are expected in the near
future (see also Refs. \cite{and97,kps}).
Thus it is worthwhile to extend the discussion of hydrodynamic
modes in Refs.~\cite{gri97,zgn} to include damping.

In the present paper, we do this by following the standard
Chapman-Enskog procedure, as first generalized for quantum gases
by Uehling and Uhlenbeck \cite{uu33,ueh34}.
In Section II, we derive the hydrodynamic equations of a trapped Bose gas
above $T_{\rm BEC}$ which include damping due to shear viscosity
$\eta$ and thermal conductivity $\kappa$.
We also obtain integral equations which give expressions
for the transport coefficients $\eta$ and $\kappa$, and in Section III
we solve for these.
Our approximation corresponds to a lowest-order polynomial approximation
of the Chapman-Enskog results such as obtained by Uehling \cite{ueh34}.
The transport coefficients have a position dependence when we include
the effect of quantum degeneracy of the trapped Bose gas.

In Section IV, we use our results to discuss the hydrodynamic damping
of the surface normal modes in a trapped Bose gas discussed in
Ref.~\cite{gri97}.
In the low density tail of a trapped gas, the hydrodynamic description
ceases to be valid.
It is crucial to include a cutoff in the linearized hydrodynamic equations
to take this into account, as pointed out by Kavoulakis, Petthick and Smith
\cite{kps,kps2}.
We also explicitly show that with this spatial cutoff, the damping of 
normal modes given by our linearized hydrodynamic equations agrees
with the expression on which Ref.~\cite{kps} is based.

In Section V, we extend these results to the Bose-condensed region just
below $T_{\rm BEC}$.
Zaremba, Griffin and Nikuni \cite{zgn} have recently given an explicit
closed  derivation of two-fluid hydrodynamic equations starting from
a microscopic model of a trapped weakly-interacting Bose-condensed gas.
We generalize the results of Ref.~\cite{zgn} to include hydrodynamic damping.
Finally in Section VI, we give some concluding remarks.

\section{Chapman-Enskog method}\label{sect2}

In Section II-IV, we limit ourselves to a
non-condensed Bose gas above $T_{\rm BEC}$, as in Ref.~\cite{gri97}.
The atoms are described by the semi-classical kinetic equation for the
distribution function $f({\bf r},{\bf p},t)$ of a Bose gas \cite{kb}
\begin{equation}
\left [{\partial\over \partial t} + {{\bf p} \over m}\cdot
\bbox{\nabla}_r
- \bbox{\nabla} U({\bf r},t)\cdot\bbox{\nabla}_p\right ] 
f({\bf r},{\bf p},t) = 
\left. {\partial
f\over \partial t}\right\vert_{coll}.
\label{eq1}
\end{equation}

\noindent Here $U({\bf r},t)=U_0({\bf r})+2gn({\bf r},t)$ includes
an external potential $U_0({\bf r})$ as well as the Hartree-Fock (HF)
self-consistent mean field $2gn({\bf r},t)$, where $n({\bf r},t)$ is
the local density.
In usual discussions \cite{uu33,ueh34}, the HF field is omitted but it 
will play a crucial role when we discuss a Bose-condensed gas below
$T_{\rm BEC}$ (see Section V).
The quantum collision integral in the right hand side of (\ref{eq1})
is given by \cite{kb}
\begin{eqnarray}
\left. {\partial
f\over \partial t}\right\vert_{coll}&=&
4\pi g^2 \int \frac{d{\bf p}_2}{(2\pi)^3}
\int \frac{d{\bf p}_3}{(2\pi)^3}\int \frac{d{\bf p}_4}{(2\pi)^3}
\delta({\bf p}+{\bf p}_2-{\bf p}_3-{\bf p}_4)\delta\left(\frac{p^2}{2m}
+\frac{p_2^2}{2m}-\frac{p_3^2}{2m}-\frac{p_4^2}{2m}\right) \cr
&&\times[(1+f)(1+f_2)f_3f_4-ff_2(1+f_3)(1+f_4)],
\label{eq2}
\end{eqnarray}
where $f\equiv f({\bf r},{\bf p},t),f_i\equiv f({\bf r},{\bf p}_i,t)$.
The interaction strength $g=4\pi\hbar^2a/m$ is determined by the
$s$-wave scattering length $a$.

The conservation laws are obtained by multiplying (1) by 
$1$, ${\bf p}$ and $p^2$ and integrating over ${\bf p}$.
In all three cases, the integrals of the collision term in (\ref{eq2}) vanish
and one finds the general hydrodynamic equations:
\begin{mathletters}
\label{eq3}
\begin{eqnarray}
&&\frac{\partial n}{\partial t}+\bbox{\nabla} \cdot(n{\bf v})=0,
\label{eq3a}\\
&&mn\left(\frac{\partial}{\partial t}+{\bf v}\cdot \bbox{\nabla}
 \right)v_{\mu}
+\frac{\partial P_{\mu\nu}}{\partial x_{\nu}}+n\frac{\partial U}
{\partial x_{\mu}}=0, 
\label{eq3b}\\
&&\frac{\partial \varepsilon}{\partial t}+\bbox{\nabla}
\cdot(\varepsilon{\bf v})
+\bbox{\nabla}\cdot{\bf Q}+D_{\mu\nu}P_{\mu\nu}=0,
\label{eq3c}
\end{eqnarray}
\end{mathletters}

\noindent
where we have defined the usual quantities:
\begin{eqnarray}
{\rm density:}\ \ \ \ \ \ \ \ \ \ 
&&n({\bf r},t)=\int\frac{d{\bf p}}{(2\pi)^3}f({\bf r},{\bf p},t),
\label{eq4} \\
{\rm velocity:}\ \ \ \ \ \ \ \ \ \ 
&&n({\bf r},t){\bf v}({\bf r},t)
=\int\frac{d{\bf p}}{(2\pi)^3}\frac{{\bf p}}{m}f({\bf r},{\bf p},t),
\label{eq5} \\
{\rm pressure \ tensor:}\ \ \ \ \ \ \ \ \ \  
&&P_{\mu\nu}({\bf r},t)
=m\int\frac{d{\bf p}}{(2\pi)^3}\left(\frac{p_{\mu}}{m}-v_{\mu} \right)
\left(\frac{p_{\nu}}{m}-v_{\nu}\right)
f({\bf r},{\bf p},t),
\label{eq6} \\
{\rm energy\ density:}\ \ \ \ \ \ \ \ \ \ 
&&\varepsilon({\bf r},t)
=\int\frac{d{\bf p}}{(2\pi)^3}\frac{1}{2m}({\bf p}-m{\bf v})^2
f({\bf r},{\bf p},t),
\label{eq7} \\
{\rm heat\ current:}\ \ \ \ \ \ \ \ \ \ 
&&{\bf Q}({\bf r},t)
=\int\frac{d{\bf p}}{(2\pi)^3}\frac{1}{2m}({\bf p}-m{\bf v})^2
\left(\frac{{\bf p}}{m}-{\bf v}\right)
f({\bf r},{\bf p},t),
\label{eq8} \\
\mbox{\rm rate-of-strain tensor:}\ \ \ \ \ \ \ \ \ \  
&&D_{\mu\nu}=\frac{1}{2}\left(\frac{\partial v_{\mu}}{\partial x_{\nu}}+
\frac{\partial v_{\nu}}{\partial x_{\mu}}\right).
\label{eq9}
\end{eqnarray}

The first approximation for the distribution function is the 
local equilibrium form
\begin{equation}
f^{(0)}({\bf r},{\bf p},t)
 =\{ {\rm exp} [\beta({\bf p}-m{\bf v})^2/2m+U-\mu]-1 \}^{-1},
\label{eq10}
\end{equation}
where the thermodynamic variables $\beta,{\bf v}$ and $\mu$ all
depend on ${\bf r}$ and $t$, and $U({\bf r},t)$ has been defined
after (\ref{eq1}).
This expression ensures that the collision integral (\ref{eq2}) vanishes.
This gives the lowest-order approximation in $l/L$, where $l$ is the mean free
path and $L$ is a characteristic wavelength.
If one uses (\ref{eq10}) to calculate the quantities from (\ref{eq4})
to (\ref{eq8}), 
one finds that ${\bf Q}=0$ and
\begin{equation}
n({\bf r},t)=\frac{1}{\Lambda^3}g_{3/2}(z({\bf r},t)),
\label{eq11}
\end{equation}
\begin{equation}
P_{\mu\nu}({\bf r},t)=\delta_{\mu\nu}P({\bf r},t),\ \ 
P({\bf r},t)=\frac{k_{\rm B}T}{\Lambda^3}g_{5/2}(z({\bf r},t))=\frac{2}{3}
\varepsilon({\bf r},t).
\label{eq12}
\end{equation}
Here $z({\rm r},t)\equiv e^{\beta({\bf r},t)[\mu({\bf r},t)-U({\bf r},t)]}$
is the local fugacity,
$\Lambda({\bf r},t)=[2\pi\hbar^2/mk_{\rm B}T({\bf r},t)]^{1/2}$
is the local thermal de Broglie wavelength
and $g_n(z)=\sum_{l=1}^{\infty}z^l/l^n$ are the well-known
Bose-Einstein functions.
The equilibrium value of the fugacity is given by $z_0({\bf r})=e^{\beta_0
(\mu-U({\bf r}))}$, with $U({\bf r})=U_0({\bf r})+2gn_0({\bf r})$.
Putting ${\bf Q}=0$ and using (\ref{eq12}) in 
Eqs.~(\ref{eq3}), one obtains
\begin{mathletters}
\label{eq13}
\begin{eqnarray}
{\partial n\over\partial t}
+\bbox{\nabla}\cdot(n {\bf v}) &=& 0, \label{eq13a} \\
m n \left [ {\partial {\bf v}\over \partial t} 
+ ({\bf v}\cdot \bbox{\nabla}){\bf v} \right ] &=& -\
\bbox{\nabla} P 
-\ n\bbox{\nabla} U, \label{eq13b} \\
{\partial \varepsilon \over \partial t} + {5\over
3}\bbox{\nabla}\cdot (\varepsilon{\bf v})  &=&
\ {\bf v}\cdot\bbox{\nabla} P.
\label{eq13c}
\end{eqnarray}
\end{mathletters}

\noindent
The linearization of the equations in (\ref{eq13}) around equilibrium leads to
the equations (4-6) in Ref.\cite{gri97} if we ignore the HF field in
$U({\bf r},t)$.

Solutions of these equations describe undamped oscillations,
some example of which are discussed in Ref. \cite{gri97} (see also
Section IV).
In order to obtain damping of the oscillations,
we have to consider the deviation of the distribution function from the 
local equilibrium form (10).
One assumes a solution of the quantum Boltzmann equation (1) of the form
\cite{uu33,kd}:
\begin{equation}
f({\bf r},{\bf p},t)=f^{(0)}({\bf r},{\bf p},t)+
f^{(0)}({\bf r},{\bf p},t)[1+f^{(0)}({\bf r},{\bf p},t)]\psi({\bf r},
{\bf p},t),
\label{eq14}
\end{equation}
where $\psi$ expresses a small deviation from local equilibrium.
To first order in $\psi$, we can reduce the collision integral in (\ref{eq2})
to
\begin{eqnarray}
&&4\pi g^2\int \frac{d {\bf p}_2}{(2\pi)^3}
\int \frac{d {\bf p}_3}{(2\pi)^3}\int d{\bf p}_4
\delta({\bf p}+{\bf p}_2-{\bf p}_3-{\bf p}_4)\delta\left(\frac{p^2}{2m}+
\frac{p_2^2}{2m}-\frac{p_3^2}{2m}-\frac{p_4^2}{2m}\right) \cr
&&\times f^{(0)}f^{(0)}_2(1+f^{(0)}_3)(1+f^{(0)}_4)
(\psi_3+\psi_4-\psi_2-\psi)\equiv\hat L[\psi],
\label{eq16}
\end{eqnarray}
where $\psi_i\equiv\psi({\bf r},{\bf p}_i,t)$.
In the left hand side of (\ref{eq1}), we approximate $f$ by $f^{(0)}$.
The various derivatives of ${\bf v}({\bf r},t),\mu({\bf r},t),
T({\bf r},t)$ and $U({\bf r},t)$ with respect to ${\bf r}$ and $t$ can be
rewritten using the lowest-order hydrodynamic equations given in (\ref{eq13}).
The resulting linearized equation for $\psi$ is (for details, see Appendix)
\begin{equation}
\left\{\frac{{\bf u}\cdot \bbox{\nabla} T}{T}\left[\frac{mu^2}{2k_{\rm B}T}
-\frac{5g_{5/2}(z)}{2g_{3/2}(z)}\right]
+\frac{m}{k_{\rm B}T}D_{\mu\nu}\left(u_{\mu}u_{\nu}
-\frac{1}{3}\delta_{\mu\nu}u^2\right)\right\}
f^{(0)}(1+f^{(0)})=\hat L[\psi],
\label{eq15}
\end{equation}
where the thermal velocity ${\bf u}$ is defined by  
$m{\bf u}\equiv{\bf p}-m{\bf v}$ and the strain tensor $D_{\mu\nu}$
is defined in 
(\ref{eq9}).
The linearized collision operator $\hat L$ is defined by (\ref{eq16}).
This equation can be shown to have a unique solution for $\psi$ if we 
impose the constraints
\begin{equation}
\int d{\bf p}f^{(0)}(1+f^{(0)})\psi=
\int d{\bf p}p_{\mu}f^{(0)}(1+f^{(0)})\psi=
\int d{\bf p}p^2f^{(0)}(1+f^{(0)})\psi=0,
\label{eq17}
\end{equation}
which mean physically that $n, {\bf v}$ and $\varepsilon$ 
[see (\ref{eq4}),(\ref{eq5}) and (\ref{eq7})] are
determined only by the first term $f^{(0)}$ in (\ref{eq14}).
Since the left hand of (\ref{eq15}) and also $f^{(0)}$ only depend on
the relative thermal velocity ${\bf u}$, the function 
$\psi$ will also depend only on ${\bf u}$ (i.e., not separately on 
${\bf p}$ and ${\bf v}$).

It is convenient to introduce dimensionless velocity variables
\begin{equation}
\left(\frac{m}{2k_{\rm B}T}\right)^{1/2}{\bf u}\equiv\mbox{\boldmath $\xi$}.
\end{equation}
With these dimensionless velocity variables, (\ref{eq15}) becomes
\begin{eqnarray}
\frac{\pi^3}{8a^2mk_{\rm B}^2T^2}
\left\{\left(\frac{2k_{\rm B}T}{m}\right)^{1/2}
\frac{\bbox{\nabla} T\cdot {\mbox{\boldmath $\xi$}}
}{T}\left[\xi^2-\frac{5g_{5/2}(z)}{2g_{3/2}(z)}
\right]+2D_{\mu\nu}\left(\xi_{\mu}\xi_{\nu}-\frac{1}{3}\delta_{\mu\nu}
\xi^2\right)
\right\}& &f^{(0)}(\xi)[1+f^{(0)}(\xi)] \cr
& &=\hat L'[\psi],
\label{eq19}
\end{eqnarray}
with $f^{(0)}(\xi)=(z^{-1}e^{\xi^2}-1)^{-1}$ and
\begin{eqnarray}
\hat L'[\psi]&\equiv&\int d {\mbox{\boldmath $\xi$}}_2 \int d {\mbox{\boldmath $\xi$}}_3
 \int d{\mbox{\boldmath $\xi$}}_4
\delta({\mbox{\boldmath $\xi$}}+{\mbox{\boldmath $\xi$}}_2-
{\mbox{\boldmath$\xi$}}_3-{\mbox{\boldmath $\xi$}}_4)
\delta\left(\xi^2+\xi_2^2-\xi_3^2-\xi_4^2\right) \cr
&& \times f^{(0)}f^{(0)}_2(1+f^{(0)}_3)(1+f^{(0)}_4)
(\psi_3+\psi_4-\psi_2-\psi).
\label{eq20}
\end{eqnarray}
For a more detailed discussion of the mathematical structure of
(\ref{eq19}) and (\ref{eq20}), we refer to the treatment of the analogous
equations for classical gases (see for example, Ref.~\cite{fk}).
The most general solution of the integral equation (\ref{eq19}) is of the
form \cite{uu33}
\begin{equation}
\psi=\frac{\pi^3}{8a^2mk_{\rm B}^2T^2}
\left[\left(\frac{2k_{\rm B}T}{m}\right)^{1/2}
\frac{\bbox{\nabla} T\cdot {\mbox{\boldmath $\xi$}}}{T}A(\xi)
+2D_{\mu\nu}\left(\xi_{\mu}\xi_{\nu}-\frac{1}{3}\delta_{\mu\nu}
\xi^2\right)B(\xi)\right].
\label{eq21}
\end{equation}
The functions $A(\xi)$ and $B(\xi)$ obey the following integral equations:
\begin{mathletters}
\label{eq22}
\begin{eqnarray}
&&\hat L'\left[{\mbox{\boldmath $\xi$}}A(\xi)\right]={\mbox{\boldmath $\xi$}}
\left[\xi^2-\frac{5g_{5/2}(z)}
{2g_{3/2}(z)}\right]f^{(0)}(1+f^{(0)}), \label{eq22a}\\
&&\hat L'\left[\left(\xi_{\mu}\xi_{\nu}-\frac{1}{3}\delta_{\mu\nu}\xi^2\right)B(\xi)
\right]=\left(\xi_{\mu} \xi_{\nu}-\frac{1}{3}\delta_{\mu\nu}\xi^2\right)
f^{(0)}(1+f^{(0)}). \label{eq22b}
\end{eqnarray}
\end{mathletters}

\noindent
For (\ref{eq21}) to satisfy the constraints given in (\ref{eq17}), 
we must also have
\begin{equation}
\int d {\mbox{\boldmath $\xi$}}\xi^2A(\xi)f^{(0)}(1+f^{(0)})=0.
\end{equation}
Using a solution of the form (\ref{eq21}) in conjunction with (\ref{eq14}), 
one can calculate the heat current density ${\bf Q}$ in (\ref{eq8})
and the pressure tensor $P_{\mu\nu}$ in (\ref{eq6}).
One finds these have the form
\begin{mathletters}
\label{eq24}
\begin{eqnarray}
{\bf Q}&=&-\kappa \bbox{\nabla} T, \label{eq24a}\\
P_{\mu\nu}&=&\delta_{\mu\nu}P-2\eta \left[D_{\mu\nu}-\frac{1}{3}
({\rm Tr}D)\delta_{\mu\nu}\right] \label{eq24b},
\end{eqnarray}
\end{mathletters}

\noindent
where the last term in (\ref{eq24b}) involves the non-equilibrium stress
tensor.
The thermal conductivity $\kappa$ and the shear viscosity coefficient $\eta$
are given in terms of the functions $A(\xi)$
and $B(\xi)$: 
\begin{mathletters}
\label{eq25}
\begin{eqnarray}
\kappa &=&-\frac{k_{\rm B}}{48a^2}
\left(\frac{2k_{\rm B}T}{m}\right)^{1/2}
\int d{\mbox{\boldmath $\xi$}}\xi^4A(\xi)f^{(0)}(1+f^{(0)}), \label{eq25a}\\
\eta &=&-\frac{m}{120a^2}
\left(\frac{2k_{\rm B}T}{m}\right)^{1/2} 
\int d{\mbox{\boldmath $\xi$}}\xi^4B(\xi)f^{(0)}(1+f^{(0)}). \label{eq25b}
\end{eqnarray}
\end{mathletters}

Introducing (\ref{eq24}) into the general hydrodynamic equations (\ref{eq3}),
one obtains the following hydrodynamic equations with 
the effect of shear viscosity and heat conduction included:
\begin{mathletters}
\label{eq26}
\begin{eqnarray}
&&\frac{\partial n}{\partial t}+\bbox{\nabla}\cdot(n{\bf v})=0,
\label{eq26a}\\
&&mn\left(\frac{\partial}{\partial t}+{\bf v}\cdot \bbox{\nabla}\right)v_{\mu}
+\frac{\partial P}{\partial x_{\mu}}+n\frac{\partial U}{\partial x_{\mu}}=
\frac{\partial}{\partial x_{\nu}}\left\{2\eta\left[D_{\mu\nu}-\frac{1}{3}
({\rm Tr}D)\delta_{\mu\nu} \right]\right\},
\label{eq26b}\\
&&\frac{\partial \varepsilon}{\partial t}+\bbox{\nabla}
\cdot(\varepsilon{\bf v})
+(\bbox{\nabla}\cdot{\bf v})P=\bbox{\nabla}\cdot(\kappa\bbox{\nabla} T)
+2\eta\left[ D_{\mu\nu}-\frac{1}{3}({\rm Tr}D)\delta_{\mu\nu}\right]^2.
\label{eq26c}
\end{eqnarray}
\end{mathletters}

\noindent
We recall that $n,{\bf v}$ and $\varepsilon$ are still given by (4)-(7)
with $f=f^{(0)}$, which means that the expressions in (11) and (12) for
$n,\varepsilon$ and $P$ are valid.
The form of these equations can be shown to agree with those originally
obtained by Uehling and Uhlenbeck \cite{uu33}.
We note that $\eta$ and $\kappa$ are position-dependent, but only through
their dependence on the equilibrium value of the fugacity $z=z_0({\bf r})$.
One slight generalization we have made over the derivation in Ref.~\cite{uu33}
is that we have included the Hartree-Fock mean field.

\section{The transport coefficients}\label{sect5}

\subsection{The thermal conductivity}
To find the thermal conductivity as given by (\ref{eq25a}), we can introduce a
simple ansatz for the form of the function $A(\xi)$ \cite{ueh34,fk}:
\begin{equation}
A(\xi)=A\left[\xi^2-\frac{5g_{5/2}(z)}{2g_{3/2}(z)}\right].
\label{eq27}
\end{equation} 
The constant $A$ is determined by multiplying (\ref{eq22a}) by 
${\mbox{\boldmath $\xi$}}[\xi^2-5g_{5/2}(z)/2g_{3/2}(z)]$ and integrating
over $\mbox{\boldmath $\xi$}$:
\begin{eqnarray}
A&=&
\int d{\mbox{\boldmath $\xi$}} \xi^2
\left[\xi^2-\frac{5g_{5/2}(z)}{2g_{3/2}(z)}\right]^2
f^{(0)}(1+f^{(0)}) \cr
& &
\times
\left\{\int d{\mbox {\boldmath $\xi$}}
\left(\xi^2-\frac{5g_{5/2}(z)}{2g_{3/2}(z)}
\right){\mbox {\boldmath $\xi$}}
\cdot \hat L'\left[
\left(\xi^2-\frac{5g_{5/2}(z)}{2g_{3/2}(z)}
\right){\mbox {\boldmath $\xi$}}\right]\right\}^{-1} \cr
&=&\frac{15\pi^{3/2}}{4I_A}
\left[\frac{7}{2}g_{7/2}(z)-\frac{5g^2_{5/2}(z)}{2g_{3/2}(z)}\right],
\end{eqnarray}
where the integral $I_A$ is defined by
\begin{equation}
I_A\equiv
\int d{\mbox{\boldmath $\xi$}}
 \left[\xi^2-\frac{5g_{5/2}(z)}{2g_{3/2}(z)}\right]
{\mbox{\boldmath $\xi$}}
\cdot \hat L'\left[\left( \xi^2-\frac{5g_{5/2}(z)}{2g_{3/2}(z)} \right)
{\mbox{\boldmath $\xi$}}\right]
=
\int d{\mbox{\boldmath $\xi$}}
 \xi^2{\mbox{\boldmath $\xi$}}
\cdot \hat L'[\xi^2{\mbox{\boldmath $\xi$}}].
\label{eq29}
\end{equation}

\noindent
In order to evaluate the integral in (\ref{eq29}), 
it is convenient to introduce the change of variables
\begin{eqnarray}
{\mbox{\boldmath $\xi$}}&=&\frac{1}{\sqrt 2}({\mbox{\boldmath $\xi$}}_0
+{\mbox{\boldmath $\xi$}}'),\ 
{\mbox{\boldmath $\xi$}}_2=\frac{1}{\sqrt 2}({\mbox{\boldmath $\xi$}}_0
-{\mbox{\boldmath $\xi$}}'), \cr
{\mbox{\boldmath $\xi$}}_3&=&\frac{1}{\sqrt 2}({\mbox{\boldmath $\xi$}}_0
+{\mbox{\boldmath $\xi$}}''),\ 
{\mbox{\boldmath $\xi$}}_4=\frac{1}{\sqrt 2}({\mbox{\boldmath $\xi$}}_0
-{\mbox{\boldmath $\xi$}}'').
\end{eqnarray}

\noindent
Then we introduce transformations from $\xi'_x\xi'_y\xi'_z$
to $\xi'\theta'\phi'$ and $\xi''_x\xi''_y\xi''_z$ to
$\xi''\theta''\phi''$, where $\theta',\theta''$ and $\phi',\phi''$ 
are the polar and azimuthal angles
with respect to the vector ${\mbox{\boldmath$\xi$}}_0$.
One obtains the following expression for $I_A$,
\begin{eqnarray}
I_A&=&-4\sqrt{2}\pi^3 I_A', \cr
I_A'(z)&=&\int_0^{\infty} d\xi_0\int_0^{\infty}d\xi'
\int_0^1dy'\int_0^1dy''\xi_0^4{\xi'}^7F(\xi_0,\xi',y',y'';z)
[{y'}^2+{y''}^2-2{y'}^2{y''}^2],
\label{eq31}
\end{eqnarray}
where $y'=\cos \theta',y''=\cos \theta''$. Here the function $F$ is 
defined by
\begin{equation}
F\equiv f^{(0)}f^{(0)}_2(1+f^{(0)}_3)(1+f^{(0)}_4)=
\frac{z^2e^{-(\xi_0^2+{\xi'}^2})}
{(1-ze^{-\xi^2})(1-ze^{-\xi_2^2})(1-ze^{-\xi_3^2})(1-ze^{-\xi_4^2})},
\label{eq32}
\end{equation}
with
\begin{eqnarray}
\xi^2&=&\frac{1}{2}(\xi_0^2+2\xi_0\xi'y'+{\xi'}^2),\ 
\xi_2^2=\frac{1}{2}(\xi_0^2-2\xi_0\xi'y'+{\xi'}^2), \cr
\xi_3^2&=&\frac{1}{2}(\xi_0^2+2\xi_0\xi'y''+{\xi'}^2),\ 
\xi_4^2=\frac{1}{2}(\xi_0^2-2\xi_0\xi'y''+{\xi'}^2).
\end{eqnarray}

Inserting the expression in (\ref{eq27}) into (\ref{eq25a}) 
and carrying out the integration,
we obtain the following expression for the thermal conductivity $\kappa$:
\begin{equation}
\kappa=-\frac{75k_{\rm B}}{64a^2m}\left(\frac{mk_{\rm B}T}{\pi}\right)^{1/2}
\frac{\pi^{1/2}}{16I'_A(z)}
\left[\frac{7}{2}g_{7/2}(z)-\frac{5g^2_{5/2}(z)}{2g_{3/2}(z)} \right]^2,
\label{eq34}
\end{equation}
where the function $I_A'(z)$ is defined in (\ref{eq31}).

\subsection{The shear viscosity}
In evaluating the shear viscosity in (\ref{eq25b}), the simplest consistent 
approximation \cite{ueh34,fk} is to use $B(\xi)\equiv B$.
The constant $B$ can be determined by multiplying (\ref{eq22b}) by
$(\xi_{\mu}\xi_{\nu}-\delta_{\mu\nu}\xi^2/3)$ and integrating over $\xi$,
\begin{eqnarray}
B&=&
\left\{\int d{\mbox{\boldmath $\xi$}}\left(\xi_{\mu}\xi_{\nu}
-\frac{1}{3}\delta_{\mu\nu}\xi^2\right)^2
f^{(0)}(1+f^{(0)})\right\}
\left\{\int d{\mbox{\boldmath $\xi$}}
\left(\xi_{\mu}\xi_{\nu}-\frac{1}{3}\delta_{\mu\nu}\xi^2\right)
\hat L'\left[\xi_{\mu}\xi_{\nu}
-\frac{1}{3}\delta_{\mu\nu}\xi^2 \right]\right\}^{-1}
 \cr
&=&
\frac{5\pi^{3/2}g_{5/2}(z)}{2I_B}.
\label{eq35}
\end{eqnarray}
The function $I_B$ is defined by
\begin{eqnarray}
I_B&=&
\int d{\mbox{\boldmath$\xi$}} 
\left(\xi_{\mu}\xi_{\nu}-\frac{1}{3}\delta_{\mu\nu}\xi^2\right)
\hat L'\left[\xi_{\mu}\xi_{\nu}
-\frac{1}{3}\delta_{\mu\nu}\xi^2 \right]
=
\int d{\mbox{\boldmath$\xi$}} 
(\xi_{\mu}\xi_{\nu})\hat L'[\xi_{\mu}\xi_{\nu}] \cr
&\equiv&-2\sqrt 2\pi^3 I'_B,
\end{eqnarray}
where
\begin{equation}
I'_B(z)=\int_0^{\infty}d\xi_0\int_0^{\infty}d\xi'\int_0^1dy'\int_0^1dy''
F(\xi_0,\xi',y',y'';z)\xi_0^2{\xi'}^7(1+y'^2+y''^2-3y'^2y''^2).
\label{eq37}
\end{equation}
This involves the same function $F$ as defined in (\ref{eq32}).
Using the ansatz $B(\xi)\equiv B$ in conjunction with 
(\ref{eq35}) in (\ref{eq25b}), we can carry out the integral.
Our final the expression for the viscosity $\eta$ is
\begin{equation}
\eta=\frac{5}{16}\frac{1}{a^2}\left(\frac{mk_{\rm B}T}{\pi}\right)^{1/2}
\frac{\pi^{1/2}}{8I'_B(z)}g^2_{5/2}(z).
\label{eq38}
\end{equation}

\subsection{High-temperature limit}
The formulas in (\ref{eq34}) and (\ref{eq38}) give $\kappa$ and $\eta$ 
as a function of the fugacity $z$. 
We note that the integrals $I_A'$ and $I_B'$ also
depend on $z$. 
These four-dimensional integrals in (\ref{eq31}) and (\ref{eq37}) can be
evaluated numerically.
In Figs.~1 and 2, we plot both $\kappa$ and $\eta$ 
as a function of the fugacity $z$.
These graphs are valid for any trapping potential 
$U_0({\bf r})$ since the latter only enters into the equilibrium fugacity 
$z_0=\exp[\beta_0(\mu-U_0({\bf r})-2gn_0({\bf r}))]$.

In the high-temperature limit, where the fugacity $
z=e^{\beta(\mu-U({\bf r}))}$
is small,
the local distribution function $f^{(0)}$ can be expanded in terms of $z$.
To third order in $z$, the function $F$ in (\ref{eq32}) reduces to
\begin{equation}
F=z^2e^{-(\xi_0^2+\xi'^2)}+2z^3e^{-\frac{3}{2}(\xi_0^2+\xi'^2)}
[\cosh(\xi_0\xi'y')+\cosh(\xi_0\xi'y'')]+O(z^4).
\end{equation}
The integrals in (\ref{eq31}) and (\ref{eq37}) can be evaluated analytically,
\begin{equation}
I'_B(z)=2I'_A(z)=\pi^{1/2}z^2\left(1+z\frac{9}{16}\sqrt{\frac{3}{2}}\right).
\end{equation}
We thus obtain the following explicit expressions for the transport
coefficients to first order in the fugacity $z$:
\begin{mathletters}
\label{eq41}
\begin{eqnarray}
\kappa&=&
\frac{1}{8}\left(\frac{75}{64}\right)\frac{k_{\rm B}}{a^2}
\left(\frac{k_{\rm B}T}{\pi m}
 \right)^{1/2}\left[1+z\left(\frac{7\sqrt{2}}{16}-\frac{9}{16}\sqrt{\frac
{3}{2}}\right)\right] \cr
&\approx&\frac{1}{8}\left(\frac{75}{64}\right)\frac{k_{\rm B}}{a^2}
\left(\frac{k_{\rm B}T}{\pi m}
 \right)^{1/2}(1-0.07n_0\Lambda^3), \label{eq41a}\\
\eta&=&\frac{1}{8}\left(\frac{5}{16}\right)\frac{m}{a^2}
\left(\frac{k_{\rm B}T}{\pi m}\right)^{1/2}
\left[1+z\left(\frac{1}{2\sqrt{2}}-\frac{9}{16}\sqrt{\frac{3}{2}}\right
)\right] \cr
&\approx&\frac{1}{8}\left(\frac{5}{16}\right)\frac{m}{a^2}
\left(\frac{k_{\rm B}T}{\pi m}\right)
^{1/2}\left(1-0.335n_0\Lambda^3\right) \label{eq41b}.
\end{eqnarray}
\end{mathletters}

\noindent
The local equilibrium density $n_0({\bf r})$ in the high-temperature limit is 
the classical result $n_0=\Lambda^{-3}
e^{\beta (\mu-U({\bf r}))}$.
The terms first order in $z$ in (\ref{eq41a}) and (\ref{eq41b})
give the first order corrections to the classical results due to Bose
statistics.
If we ignore the HF mean field $2gn_0({\bf r})$,
these transport coefficients reduce to the expressions first obtained
for a uniform Bose gas by Uehling \cite{ueh34}.
The Bose quantum corrections to the classical results in both $\eta$ and 
$\kappa$ depend on the local fugacity $z$ and, through this, on position
in a trapped gas.

The density-independent terms in Eqs. (\ref{eq41a}) and 
(\ref{eq41b}) are 8 times smaller than
the well-known Chapman-Enskog expressions for classical hard spheres.
This is due to the difference (see also Ref. \cite{ueh34})
in the quantum binary scattering cross-section
for Bosons when correctly calculated using symmetrized wavefunctions
($\sigma=8\pi a^2$, instead of $\pi a^2$).

\section{Hydrodynamic damping of normal modes}
The linearized version of the hydrodynamic equations in (\ref{eq26}) are
\begin{mathletters}
\label{eq42}
\begin{eqnarray}
&&\frac{\partial \delta n}{\partial t}+\bbox{\nabla}
 \cdot(n_0\delta{\bf v})=0, 
\label{eq42a} \\
&&mn_0\frac{\partial \delta v_{\mu}}{\partial t}
=-\frac{\partial \delta P}{\partial x_{\mu}} 
- \delta n \frac{\partial U}{\partial x_{\mu}} 
-2gn_0\frac{\partial \delta n}{\partial x_{\mu}} +
\frac{\partial}{\partial x_{\nu}}\left\{2\eta\left[D_{\mu\nu}-\frac{1}{3}
({\rm Tr}D)\delta_{\mu\nu} \right]\right\},
\label{eq42b} \\
&&\frac{\partial \delta P}{\partial t}=
-\frac{5}{3}\bbox{\nabla}\cdot(P_0\delta {\bf v})+\frac{2}{3}\delta{\bf v}
\cdot\bbox{\nabla} P_0+\frac{2}{3}\bbox{\nabla}\cdot(\kappa\bbox{\nabla} T),
\label{eq42c}
\end{eqnarray}
\end{mathletters}

\noindent
where repeated Greek subscripts are summed over.

\noindent
Taking the time derivative of (\ref{eq42b}) and using (\ref{eq42a}) and 
(\ref{eq42c}) gives
an equation for the velocity fluctuations which is only coupled to
the temperature fluctuations:
\begin{eqnarray}
m\frac{\partial^2 \delta v_{\mu}}{\partial t^2}&=&
\frac{5P_0}{3n_0}\frac{\partial}{\partial x_{\mu}}
(\bbox{\nabla}\cdot\delta{\bf v})-
\frac{\partial}{\partial x_{\mu}}(\delta{\bf v}
\cdot\bbox{\nabla} U_0)-\frac{2}{3}(\bbox{\nabla}
\cdot\delta{\bf v})\frac{\partial U_0}{\partial x_{\mu}} \cr
&&+2g\frac{\partial}{\partial x_{\mu}}[n_0(\bbox{\nabla}\cdot\delta{\bf v})]
-\frac{4}{3}g(\bbox{\nabla}\cdot\delta{\bf v})\frac{\partial}{\partial x_{\mu}}
n_0 \cr
&&+\frac{1}{n_0}\frac{\partial}{\partial x_{\nu}}
\left\{2\eta\left[\frac{\partial}{\partial t}D_{\mu\nu}
-\frac{1}{3}\left({\rm Tr}\frac{\partial D}{\partial t}\right)
\delta_{\mu\nu}\right]\right\}
-\frac{2}{3n_0}\frac{\partial}{\partial x_{\mu}}\bbox{\nabla}\cdot
\left(\kappa\bbox{\nabla}\delta T\right).
\label{eq44}
\end{eqnarray}
Here we have used the relation $\bbox{\nabla} P_0=-n_0\bbox{\nabla} U$
valid in equilibrium.
Since we assume the dissipative terms to be small,
we can use the lowest-order hydrodynamic equation for $\delta T$ in 
(\ref{eq44}) (see (A7) in Appendix), 
\begin{equation}
\frac{\partial \delta T}{\partial t}=-\frac{2}{3}T_0\bbox{\nabla}\cdot
\delta{\bf v}.
\label{eq45}
\end{equation}
Equations (\ref{eq44}) and (\ref{eq45}) are thus a closed set of equations
for the fluctuations.
Equation (\ref{eq44}) generalizes that given by (13) in Ref.\cite{gri97}
to include damping due to viscosity and thermal conductivity as well as the
HF mean field $2gn$.

At this point, we introduce a crucial 
aspect of a trapped gas which leads to an important modification of the
preceding hydrodynamic equations.
Kavoulakis et al.~\cite{kps} have
pointed out the breakdown of the local equilibrium
(hydrodynamic) description in the low density outer region of a trapped gas.
The corrections involving thermal conduction and viscous processes are
expected to vanish (over a mean free path) in the dilute region where
collisions become ineffective in producing local equilibrium. 
This effect can be simulated in the hydrodynamic equations by
multiplying the hydrodynamic expression for the thermal conductivity
in (\ref{eq24a}) and the viscous stress tensor
in (\ref{eq24b}) by a step function that vanishes outside the 
cloud \cite{kps2}:
\begin{mathletters}
\label{neweq1}
\begin{eqnarray}
{\bf Q}&=&-\kappa\bbox{\nabla}T\Theta(r_0-r),\\
\label{neweq1a}
P_{\mu\nu}&=&\delta_{\mu\nu}P-2\eta\left[D_{\mu\nu}-\frac{1}{3}(TrD)
\delta_{\mu\nu}\right]\Theta(r_0-r),
\label{neqeq1b}
\end{eqnarray}
\end{mathletters}

\noindent
For simplicity of notation, we only consider an isotropic trap, in which
case the hydrodynamic description breaks down on a sphere of radius $r_0$. 
A method of calculating $r_0$ is given in Ref.~\cite{kps} for a classical
trapped gas.
The equation of motion in (\ref{eq44}) for the velocity fluctuations is
modified to
\begin{eqnarray}
m\frac{\partial^2 \delta v_{\mu}}{\partial t^2}&=&
\frac{5P_0}{3n_0}\frac{\partial}{\partial x_{\mu}}
(\bbox{\nabla}\cdot\delta{\bf v})-
\frac{\partial}{\partial x_{\mu}}(\delta{\bf v}
\cdot\bbox{\nabla} U_0)-\frac{2}{3}(\bbox{\nabla}\cdot\delta{\bf v})
\frac{\partial U_0}{\partial x_{\mu}} \cr
&&+2g\frac{\partial}{\partial x_{\mu}}[n_0(\bbox{\nabla}\cdot\delta{\bf v})]
-\frac{4}{3}g(\bbox{\nabla}\cdot\delta{\bf v})\frac{\partial n_0}
{\partial x_{\mu}} \cr
&&+\frac{1}{n_0}
\frac{\partial}{\partial x_{\nu}}
\left\{2\eta\left[\frac{\partial}{\partial t}D_{\mu\nu}
-\frac{1}{3}\left({\rm Tr}\frac{\partial D}{\partial t}\right)
\delta_{\mu\nu}\right]\Theta(r_0-r)\right\}\cr
&&-\frac{2}{3n_0}\frac{\partial}{\partial x_{\mu}}\bbox{\nabla}\cdot
[\kappa\bbox{\nabla}\delta T\Theta(r_0-r)].
\label{neweq2}
\end{eqnarray}

We now look for normal mode solutions of (\ref{neweq2}) with the convention
$\delta {\bf v}({\bf r},t)=\delta{\bf v}_{\omega}({\bf r})
 e^{-i\omega t}$,
$\delta T({\bf r},t)=\delta T_{\omega}({\bf r}) e^{-i\omega t}$ and
$D({\bf r},t)=D_{\omega}({\bf r}) e^{-i\omega t}$.
Combining (\ref{eq45}) and (\ref{neweq2}), we obtain a closed equation
for $\delta {\bf v}_{\omega}$: 
\begin{eqnarray}
&&-mn_o(\omega^2\delta v_{\omega\mu}-\hat W_{\mu}[\delta{\bf v}_{\omega}]) \cr
&=&-i\omega\frac{\partial}{\partial x_{\nu}}
\left\{2\eta\left[D_{\omega\mu\nu}-\frac{1}{3}({\rm Tr}D_{\omega})
\delta_{\mu\nu}\right]\Theta(r_0-r)\right\} \cr
&&+i\frac{4T_0}{9n_0\omega}\frac{\partial}{\partial x_{\mu}}
\{\bbox{\nabla}\cdot[\kappa\bbox{\nabla}
(\bbox{\nabla}\cdot\delta{\bf v}_{\omega})\Theta(r_0-r)]\},
\label{neweq3}
\end{eqnarray}
where the operator $\hat W_{\mu}[\delta {\bf v}_{\omega}]$ is defined by
\begin{eqnarray}
-m\hat W_{\mu}[\delta {\bf v}_{\omega}]&\equiv&
\frac{5P_0}{3n_0}\frac{\partial}{\partial x_{\mu}}
(\bbox{\nabla}\cdot\delta{\bf v}_{\omega})-
\frac{\partial}{\partial x_{\mu}}(\delta{\bf v}_{\omega}
\cdot\bbox{\nabla} U_0)-\frac{2}{3}(\bbox{\nabla}\cdot\delta{\bf v}_{\omega})
\frac{\partial U_0}{\partial x_{\mu}} \cr
&&+2g\frac{\partial}{\partial x_{\mu}}[n_0(\bbox{\nabla}\cdot\delta{\bf v}
_{\omega})]
-\frac{4}{3}g(\bbox{\nabla}\cdot\delta{\bf v}_{\omega})\frac{\partial n_0}
{\partial x_{\mu}}.
\label{neweq4}
\end{eqnarray}
The undamped solution $\delta \tilde {\bf v}_{\omega}$ of (\ref{neweq3}) when
$\kappa,\eta=0$ is given by
$\hat W_{\mu}[\delta \tilde {\bf v}_{\omega}]={\tilde\omega}^2 
\delta \tilde v_{\omega\mu}$
(a tilde denotes undamped solution).
For later use, we note that the operator $\hat W_{\mu}$ has the following
property
\begin{equation}
\int n_0 \delta v^*_{\omega'\mu}\hat W_{\mu}[\delta{\bf v}_{\omega}]d{\bf r}
=\int n_0 \delta v_{\omega\mu}\hat W_{\mu}[\delta{\bf v}^*_{\omega'}]d{\bf r}.
\end{equation}
Therefore the undamped solutions $\delta \tilde {\bf v}_{\omega}$ satisfy the
following orthogonality relation
\begin{equation}
\int n_0 \delta \tilde v^*_{\omega'\mu}\hat 
W_{\mu}[\delta \tilde {\bf v}_{\omega}]d{\bf r}=0,\ \ \ {\rm if}\ 
\tilde\omega'\neq \tilde\omega.
\end{equation}.

We consider an isotropic trap $U_0({\bf r})=\frac{1}{2}m\omega_0^2r^2$.
The undamped solution for a monopole or breathing mode is described by 
$\delta \tilde{\bf v}_{\omega}({\bf r})\sim {\bf r}$,
the undamped frequency is  found to be $\tilde\omega=2\omega_0$ \cite{gri97}.
Using this solution in (\ref{neweq3}), 
we find $D_{\omega\mu\nu}\sim\delta_{\mu\nu}$ and hence 
there is no shear viscous contribution to the damping of this mode.
In addition, since $\bbox{\nabla} \cdot \delta\tilde{\bf v}_{\omega}=$const, 
the last term in (\ref{neweq3}) involving the thermal conductivity 
coefficient $\kappa$ does not contribute.
We conclude the {\it monopole mode has no damping}
for an isotropic harmonic trap.
As noted in Ref.~\cite{gri97}, this surprising result was first derived
for a classical gas by Boltzmann in 1876.

The undamped solution
for divergence-free surface modes 
$(\bbox{\nabla} \cdot \delta \tilde{\bf v}_{\omega}=0)$ is given by 
\cite{gri97} 
$\delta \tilde{\bf v}_{\omega}
({\bf r})=\bbox{\nabla}\chi_{\omega}({\bf r})$, where
\begin{equation}
\chi_{\omega}({\bf r})\propto r^lY_{lm}(\theta,\phi),
\label{eq46}
\end{equation}
with the dispersion relation $\tilde\omega=\sqrt{l} \omega_0$, $l=1,2,\cdots$.
Using the solution (\ref{eq46}) in the right hand side of (\ref{neweq3}),
we find (there is no contribution from the thermal conductivity because
$\bbox{\nabla}\cdot \delta\tilde{\bf v}_{\omega}=0$).
\begin{eqnarray}
{\rm RHS \ of \ (47)}&=&-2i\omega\frac{\partial}{\partial x_{\nu}}\left[\eta
\frac{\partial^2 \chi_{\omega}}{\partial x_{\mu}\partial x_{\nu}}
\Theta(r_0-r)\right] \cr
&=&2i\omega
\left[\frac{z}{k_{\rm B}T}\frac{\partial \eta}{\partial z}m\omega_0^2
\Theta(r_0-r)+\frac{\eta}{r}\delta(r-r_0)\right]
x_{\nu}\frac{\partial^2 \chi_{\omega}}{\partial x_{\mu}\partial x_{\nu}} \cr
&=&2i\omega(l-1)
\left[\frac{z}{k_{\rm B}T}\frac{\partial\eta}{\partial z}m\omega_0^2
\Theta(r_0-r)+\frac{\eta}{r_0}\delta(r-r_0)\right]
\delta v_{\omega\mu}.
\label{eq48}
\end{eqnarray}
For $l=1$, we see that the right hand side of (\ref{neweq3}) vanishes. 
This means that the center-of-mass mode
with frequency $\omega_0$ has no hydrodynamic damping.
For $l>1$, the undamped solution is no longer the solution of the
equation of motion (\ref{neweq3}).
In this case, Eq.~(\ref{neweq3}) must be solved for $r<r_0$ with a boundary
condition at $r=r_0$ to take into account the discontinuity at $r_0$ given
in (\ref{eq48}).

To complete this section, we derive a general expression for hydrodynamic
damping from the linearized hydrodynamic equations for the velocity
fluctuations given by (\ref{neweq3}).
Multiplying (\ref{neweq3}) by $\delta v^*_{\omega\mu}$ and integrating over 
${\bf r}$, we obtain
\begin{eqnarray}
&&-m\omega^2\int n_0|\delta v_{\omega}|^2 d{\bf r}+m\int n_0\delta v^*_{\omega\mu}
\hat W_{\mu}[\delta {\bf v}_{\omega}] d{\bf r} \cr
&=&-i{\omega}
\int \delta v^*_{\omega\mu}
\frac{\partial}{\partial x_{\nu}}
\left\{2\eta\left[D_{\omega\mu\nu}-\frac{1}{3}({\rm Tr}D_{\omega})
\delta_{\mu\nu}\right]\Theta(r_0-r)\right\} d{\bf r} \cr
&&+\frac{4T_0}{9n_0\omega}\int
\frac{\partial}{\partial x_{\mu}}\{\bbox{\nabla}\cdot[\kappa\bbox{\nabla}
(\bbox{\nabla}\cdot\delta{\bf v}_{\omega})\Theta(r_0-r)]\} d{\bf r} \cr
&=&i{\omega}
\int \left[
2\eta\left|D_{\omega\mu\nu}-\frac{1}{3}({\rm Tr}D_{\omega})
\delta_{\mu\nu}\right|^2
+\frac{4T_0}{9n_0\omega^2} \kappa |\bbox{\nabla}
(\bbox{\nabla}\cdot\delta{\bf v}_{\omega})|^2\right]
\Theta(r_0-r) d{\bf r}.
\label{neweq5}
\end{eqnarray}
Since the undamped solutions $\delta \tilde {\bf v}_{\omega}$  
satisfy the orthogonality relation (50), one sees that these undamped
solutions can be used to find the eigenvalue $\omega^2$ of (\ref{neweq5})
to first order in $\eta$ and $\kappa$.
Inserting the undamped solutions into (\ref{neweq5}), we find it reduces
\begin{eqnarray} 
&&
-m(\omega^2-\tilde\omega^2)\int n_0|\delta \tilde v_{\omega}|^2 d{\bf r} \cr
&=&i{\omega}
\int \left[
2\eta\left|\tilde D_{\omega\mu\nu}-\frac{1}{3}({\rm Tr}\tilde D_{\omega})
\delta_{\mu\nu}\right|^2
+\frac{4T_0}{9n_0\omega^2} \kappa |\bbox{\nabla}
(\bbox{\nabla}\cdot\delta\tilde{\bf v}_{\omega})|^2\right]
\Theta(r_0-r) d{\bf r}.
\label{neweq6}
\end{eqnarray}
Assuming the complex solution $\omega=\tilde\omega - i\Gamma$,
one finds the damping rate $\Gamma$ is given by the expression
(valid to first order in $\eta$ and $\kappa$)
\begin{eqnarray}
\Gamma&=&
\int \left[
2\eta\left|\tilde D_{\omega\mu\nu}-\frac{1}{3}({\rm Tr}\tilde D_{\omega})
\delta_{\mu\nu}\right|^2
+\frac{4T_0}{9n_0{\tilde \omega^2}} \kappa |\bbox{\nabla}
(\bbox{\nabla}\cdot\delta\tilde{\bf v}_{\omega})|^2\right]
\Theta(r_0-r) d{\bf r} \cr
&&\times\left[ 2m\int n_0|\delta \tilde v_{\omega}|^2d{\bf r}\right]^{-1}.
\label{neweq7}
\end{eqnarray}
The damping rate given by (\ref{neweq7}) is precisely \cite{kps2}
the one
obtained in  Ref.~\cite{kps}, where the damping rate is calculated
from the rate of increase of the total entropy \cite{ll}.
The transport coefficients $\kappa$ and $\eta$ in (\ref{neweq7}) are given
in Figs.~1 and 2 for a trapped Bose gas and in general 
are dependent on position through the fugacity.

\section{Generalization to a superfluid gas}
In this section, we extend the hydrodynamic equations derived in Section II
to the case of a superfluid trapped gas by generalizing the results of
Ref.~\cite{zgn} to include hydrodynamic damping.
The two-fluid equations in Ref.~\cite{zgn} consist of equations of motion
for the condensate and a set of hydrodynamic equations for the non-condensate.
The latter description is only valid for in the semiclassical region
$k_{\rm B}T \gg \hbar\omega_0, gn_0$.
The time-dependent Hartree-Fock-Popov equation of motion for the condensate
wavefunction can be rewritten in terms of a
pair of hydrodynamic equations for the condensate density $n_c$
and the superfluid velocity ${\bf v}_s$,
\begin{mathletters}
\label{eq55}
\begin{eqnarray}
{\partial n_c\over \partial t} &=& -\bbox{\nabla} \cdot
(n_c{\bf v}_s),  \label{eq55a}\\
m\left ({\partial {\bf v}_s\over \partial t} \right. &+&
\left. {1\over
2}\bbox{\nabla}{\bf v}_s^2 \right ) = -\bbox{\nabla}
\phi. \label{eq55b}
\end{eqnarray}
\end{mathletters}

\noindent
The potential $\phi$ is defined by \cite{zgn}
\begin{equation}
\phi({\bf r},t)\equiv-\frac{\hbar^2\nabla^2 [n_c({\bf r},t)]^{1/2}}
{2m [n_c({\bf r},t)]^{1/2}}+U_0({\bf r})+2g\tilde n({\bf r},t)+gn_c({\bf r},t),
\label{eq56}
\end{equation}
where $\tilde n({\bf r},t)$ is the non-condensate density
describing the excited atoms.
The first term in (\ref{eq56}) is sometimes called the quantum pressure term
and vanishes in uniform systems.
It is associated with the spatially varying amplitude of the Bose order
parameter.

The equations of motion for the non-condensate can be derived from the
kinetic equation (\ref{eq1}) for the distribution
function $f({\bf r},{\bf p},t)$ of the excited atoms.
The HF mean field in $U({\bf r},t)$ is now
$2gn({\bf r},t)=2g[\tilde n({\bf r},t)+n_c({\bf r},t)]$,
where $\tilde n(n_c)$ represents the local density of the non-condensate
(condensate) atoms.
Apart from this change, the expression for the collision integral
in (\ref{eq2}) is still applicable at finite temperatures
below $T_{\rm BEC}$.
Thus the hydrodynamic equations for the non-condensate
can be derived using precisely the same procedure
developed in Section II for $T>T_{\rm BEC}$.
We obtain equations analogous to (\ref{eq26}) in the following form:
\begin{mathletters}
\label{eq57}
\begin{eqnarray}
&&\frac{\partial \tilde n}{\partial t}+\bbox{\nabla}
\cdot(\tilde n{\bf v}_n)=0,
\label{eq57a}
\\
&&m\tilde n\left(\frac{\partial}{\partial t}+{\bf v}_n\cdot
 \bbox{\nabla}\right)v_{n\mu}
+\frac{\partial \tilde P}{\partial x_{\mu}}+\tilde n\frac{\partial U}
{\partial x_{\mu}}=
\frac{\partial}{\partial x_{\nu}}\left\{2\eta\left[D_{\mu\nu}-\frac{1}{3}
({\rm Tr}D)\delta_{\mu\nu} \right]\right\},
\label{eq57b}
\\
&&\frac{\partial \tilde \varepsilon}{\partial t}+\bbox{\nabla}
\cdot(\tilde \varepsilon{\bf v}_n)
+(\bbox{\nabla}\cdot{\bf v}_n)\tilde P=\bbox{\nabla}\cdot(\kappa\bbox{\nabla}
T)
+2\eta\left[ D_{\mu\nu}-\frac{1}{3}({\rm Tr}D)\delta_{\mu\nu}\right]^2.
\label{eq57c}
\end{eqnarray}
\end{mathletters}

\noindent
Here $\tilde n$, $\tilde P$ and $\tilde \varepsilon$ are given by the same
expressions as $n$, $P$ and $\varepsilon$ in (\ref{eq11}) and
(\ref{eq12}).
$D_{\mu\nu}$ is defined as in (\ref{eq9}), where now 
the velocity ${\bf v}$ is the normal fluid velocity ${\bf v}_n$.
The expressions for the transport coefficients given in 
(\ref{eq34}) and (\ref{eq38})
are directly applicable below $T_{\rm BEC}$ for any trapping potential,
apart from change in  the form of the HF field in the local fugacity $z$.
As in the case above $T_{\rm BEC}$ considered in Section IV, one must
introduce a cutoff in the linearized hydrodynamic equations to take into
account that the terms proportional to the transport coefficients
vanish when the density becomes too low to sustain local equilibrium.

We might note that the results for $\kappa$ and $\eta$ by Kirkpatrick and
Dorfman \cite{kd,footnote} for a uniform gas below $T_{\rm BEC}$ 
have an additional factor $(1+b n_c\Lambda^3)^{-1}$, where $b$ is of order 1.
This factor arises from collisions between condensate and excited atoms,
which are important at very low temperatures ($T \ll T_{\rm BEC}$).
Our present analysis ignores this contribution.
In this regard, we note that (\ref{eq1}) and (\ref{eq2}) are based on
validity of a simple semi-classical Hartree-Fock description of excited atoms
(ignoring the off-diagonal or anomalous self-energy terms).
This has been shown \cite{gps} to give an excellent approximation for the
thermodynamic properties of a trapped Bose gas down to much lower
temperatures ($T\ll T_{\rm BEC}$) than in a uniform gas.

If we neglect the first term in (\ref{eq56}), the equations in 
(\ref{eq55}) and (\ref{eq57})
can be used to derive \cite{zgn} the two-fluid  equations for a trapped
Bose-condensed gas in the Landau form \cite{khala},
except that our equations do not include any effect from the bulk
viscosities.
In a uniform superfluid Bose gas,
Kirkpatrick and Dorfman \cite{kd} found
that the bulk viscosities vanish in the finite temperature
region ($k_{\rm B}T\gg gn_0$) we have been considering.
Our present results for a trapped Bose gas are consistent with this
and moreover justify the omission of any dissipative terms in the condensate
equation of motion (\ref{eq55b}).

\section{Concluding remarks}

Summarizing our main results, starting from the microscopic equations
of motion,
we have derived a closed equation (\ref{eq44})
for the velocity fluctuations of a trapped Bose gas, which
includes damping due to hydrodynamic processes.
We have also obtained explicit expressions for the shear viscosity $\eta$
and the thermal conductivity $\kappa$ for a trapped Bose gas,
which depend on position through the local fugacity $z_0$.
We have given a detailed derivation for a trapped gas above $T_{\rm BEC}$,
and more briefly discussed the generalization of these results
to a Bose-condensed gas below $T_{\rm BEC}$ in Section V.

For illustration, in Section IV, we used the linearized hydrodynamic
equations to discuss some of
the normal modes of oscillation for a normal Bose gas above $T_{\rm BEC}$.
We show that the monopole mode of an isotropic trap was undamped.
We also note that our linearized equations in
(\ref{eq42b}) and (\ref{eq42c}) do not take into account
that in the low density region in trapped Bose gases,
the hydrodynamic description breaks down, as emphasized in 
Refs.~\cite{kps,kps2}.
We show that if a spatial cutoff is used in our linearized hydrodynamic
equations, the damping of the normal mode solutions is given by the
same expression as one finds from calculating the rate of entropy production
\cite{kps,kps2}.
In this formalism, one has reduced the calculation of the damping
to that of determining the boundary where hydrodynamics breaks down.
This has been worked out in Ref.~\cite{kps} for a classical trapped gas,
in which the transport coefficients are independent of density.
An interesting problem for the future is to use the semi-classical
kinetic equation given by (\ref{eq1}) and (\ref{eq2}) to give a more
microscopic analysis of the low density region where expanding around
the local equilibrium solution [see (14)] is no longer valid.

We note that the general form of the hydrodynamic equations given in
Eq.~(\ref{eq26}) can be derived more generally using conservation laws
for conserved quantities in conjunction with local thermal equilibrium.
This approach is developed in many textbooks \cite{cl} and can be
formally extended to deal with inhomogeneous superfluids in an external
potential (see, for example, Ref.~\cite{hs}).
This method is, of course, phenomenological in that it introduces various
linear response coefficients [relating densities (thermodynamic derivatives)
and currents (transport coefficients) to fields] whose evaluation requires
some specific microscopic model.
While very useful, this approach must be ultimately justified by a fully
microscopic treatment given by the underlying kinetic equations,
such as used in this paper and in Refs.~\cite{gri97,zgn,uu33,kd}.
In particular, as noted above, such a microscopic approach is needed to
deal with the breakdown in the local equilibrium solution in the low
density tail of trapped gases.
Moreover we note that our microscopic derivation in Section V (see also
Ref.~\cite{zgn}) gives a more complete description, as can be seen from
the fact that we obtain separate conservation laws for the condensate
and non-condensate densities.
The standard phenomenological two-fluid equations do not determine the
condensate and non-condensate density fluctuations separately, but
only the total density fluctuation. 

\acknowledgments
We would like to thank G. Kavoulakis, C. Pethick and H. Smith for a
useful correspondence concerning the approach used in Ref.~\cite{kps}
and the importance of introducing a cutoff in the linearized hydrodynamic
equations of a trapped gas.
T.N. was supported by a grant from the Japan Society for the
Promotion of Science (JSPS) and A.G. by a grant from NSERC Canada.

\appendix
\section{}
We briefly sketch the derivation of the left hand side of the kinetic
equation in (\ref{eq15}).
Using (\ref{eq10}) in the left hand side of (\ref{eq1}), one has
\begin{eqnarray}
&&\left[ \frac{\partial}{\partial t} +\frac{{\bf p}}{m}\cdot\bbox{\nabla}_r
-\bbox{\nabla}_r U({\bf r},t)\cdot\bbox{\nabla}_p\right]
f^{(0)}({\bf r},{\bf p},t) \nonumber \\
&&=\left[\frac{1}{z}\left(\frac{\partial}{\partial t}+
\frac{{\bf p}}{m}\cdot\bbox{\nabla}_r \right)z
+\frac{mv^2}{2k_{\rm B}T^2}
\left(\frac{\partial}{\partial t}+
\frac{{\bf p}}{m}\cdot\bbox{\nabla}_r \right)T \right. \nonumber \\
&&\ \ \ \  \left. +\frac{m{\bf u}}{k_{\rm B}T}\cdot
\left(\frac{\partial}{\partial t}+
\frac{{\bf p}}{m}\cdot\bbox{\nabla}_r \right){\bf v}
+\frac{\bbox{\nabla}_rU({\bf r},t)}{k_{\rm B}T}\cdot {\bf u}
\right] f^{(0)}(1+f^{(0)}).
\label{A1}
\end{eqnarray}
Using the lowest-order hydrodynamic equations in (\ref{eq13}) and the 
expressions for the density $n$ in (\ref{eq11}) and the pressure $P$ in
(\ref{eq12}), one finds
\begin{eqnarray}
\frac{\partial n}{\partial t} &=& \frac{3n}{2T}\frac{\partial T}{\partial t}
+\frac{\gamma k_{\rm B}T}{z}\frac{\partial z}{\partial t} \\
&=& -\frac{3n}{2T}{\bf v}\cdot\bbox{\nabla}T
-\frac{\gamma k_{\rm B}T}{z}{\bf v}\cdot\bbox{\nabla}z
-n\bbox{\nabla}\cdot{\bf v},
\end{eqnarray}
where $\gamma\equiv \frac{1}{k_{\rm B}T}\frac{1}{\Lambda^3}g_{1/2}(z)$,
and
\begin{eqnarray}
\frac{\partial P}{\partial t}&=&
\frac{5}{2}\frac{P}{T}
\frac{\partial T}{\partial t}+\frac{nk_{\rm B}T}{z}
\frac{\partial z}{\partial t} \\
&=&-\frac{5}{2}\frac{P}{T}{\bf v}\cdot\bbox{\nabla}T
-\frac{nk_{\rm B}T}{z}{\bf v}\cdot\bbox{\nabla}z
-\frac{5}{3}P(\bbox{\nabla}\cdot{\bf v}).
\end{eqnarray}
One may combine these equations to obtain
\begin{equation}
\frac{\partial z}{\partial t}=-{\bf v}\cdot\bbox{\nabla}z
=-\frac{z}{nk_{\rm B}T}{\bf v}\cdot\left(\bbox{\nabla}P
-\frac{5}{2}\frac{P}{T}\bbox{\nabla}T\right),
\end{equation}
\begin{equation}
\frac{\partial T}{\partial t}=-\frac{2}{3}T\bbox{\nabla}\cdot{\bf v}
-({\bf v}\cdot\bbox{\nabla})T.
\end{equation}
The analogous equation for $\partial {\bf v}/\partial t$ is given directly
by (13b).
Using these results in (A1), one finds that it reduces to
\begin{eqnarray}
&&\left(\frac{\partial}{\partial t}
+\frac{{\bf p}}{m}\cdot\bbox{\nabla}_r-\bbox{\nabla}_rU\cdot\bbox{\nabla}_p
\right)f^{(0)} \nonumber \\
&&=\left\{
\frac{1}{T}{\bf u}\cdot\bbox{\nabla}T\left(\frac{mu^2}{2k_{\rm B}T}
-\frac{5P}{2nk_{\rm B}T} \right)
+\frac{m}{k_{\rm B}T}\left[{\bf u}\cdot({\bf u}\cdot\bbox{\nabla}){\bf v}
-\frac{u^2}{3}\bbox{\nabla}\cdot{\bf v}\right]\right\}
f^{(0)}(1+f^{(0)}),
\end{eqnarray}
where we recall ${\bf u}\equiv {\bf p}/m-{\bf v}$.
This can be rewritten in the form shown on the left hand side of
(\ref{eq15}).


\vfil\break
\centerline{\bf FIGURE CAPTIONS}
\begin{itemize}
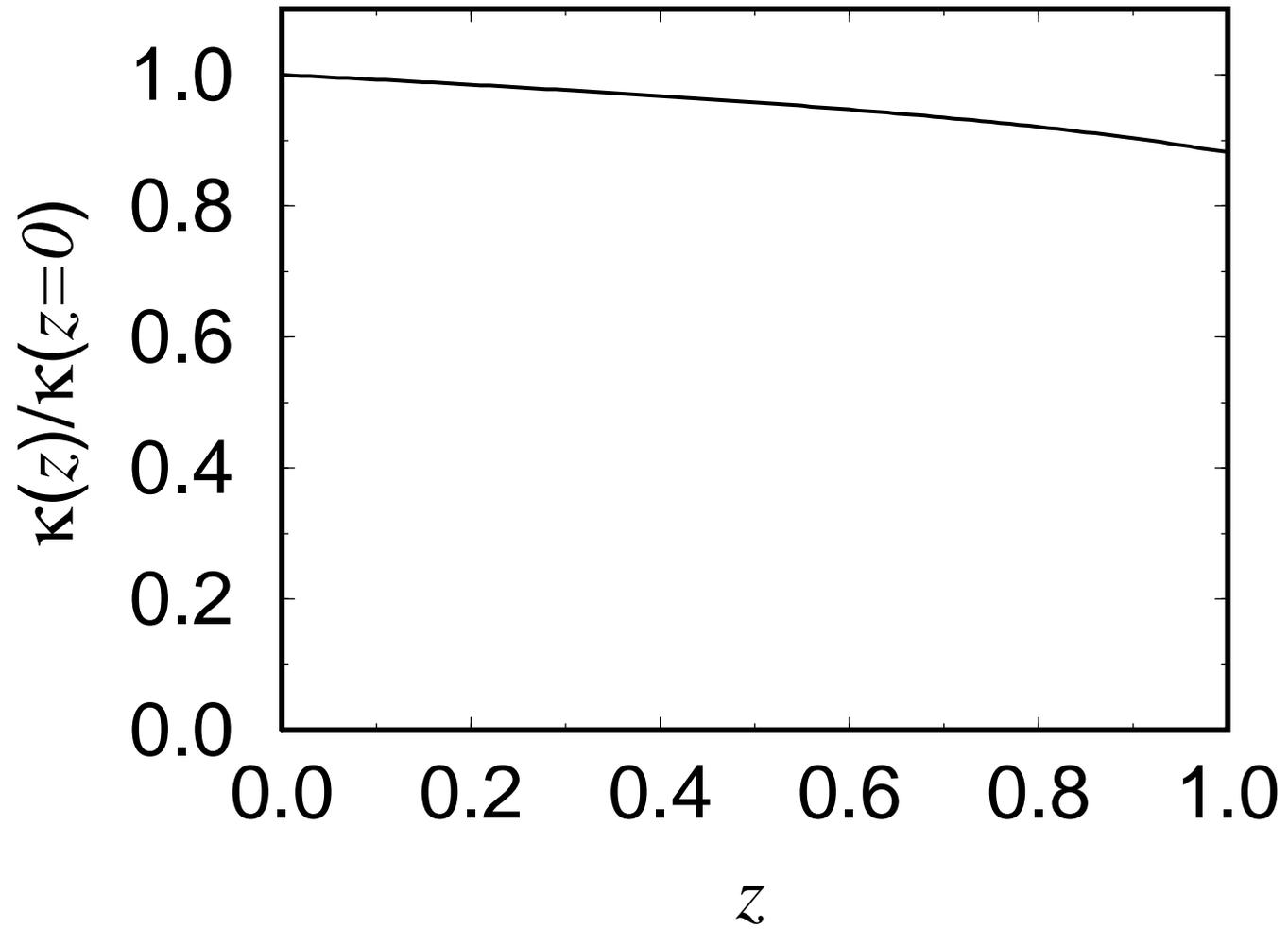
\item[Fig.1:] 
The thermal conductivity $\kappa$ as a function of 
the fugacity $z$.
The values are normalized to the classical gas results at $z=0$. 
(see (\ref{eq41a}))

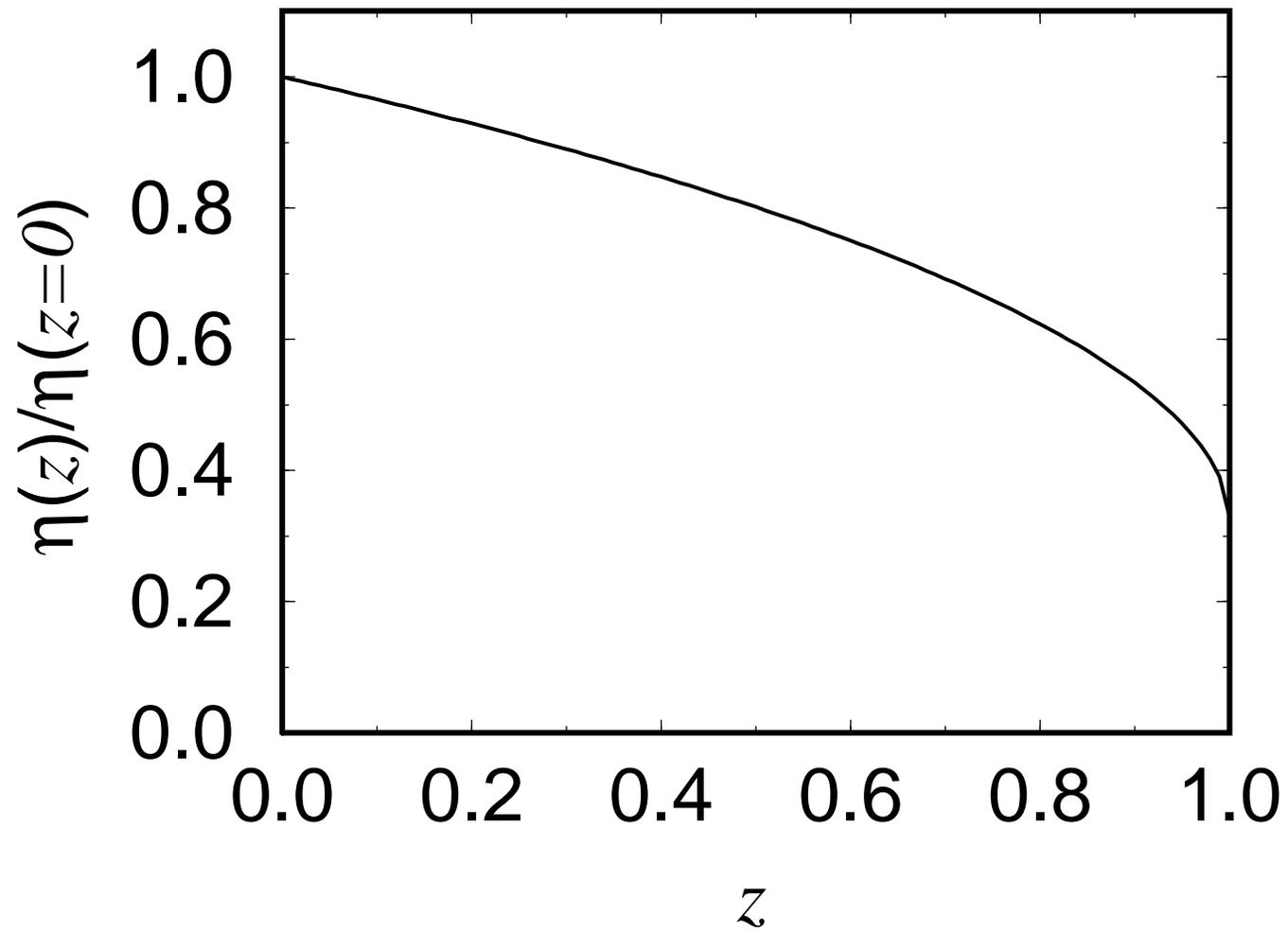
\item[Fig.2:] 
The viscosity coefficient $\eta$ as a function of
the fugacity $z$.
The values are normalized to the classical gas results at $z=0$. 
(see (\ref{eq41b}))

\end{itemize}


\begin{references}
\bibitem{mewes} D.S. Jin {\it et al.}, Phys. Rev. Lett.
{\bf 77}, 420 (1996);
M.-O.~Mewes et al., Phys.\ Rev.\ Lett., {\bf 77}, 992(1996).

\bibitem{jin97} D.S. Jin {\it et al.}, Phys. Rev. Lett.
{\bf 78}, 764 (1997).

\bibitem{gri97}A. Griffin, W.C. Wu and S. Stringari, \prl {\bf 78},
1838 (1997).

\bibitem{zgn}E. Zaremba, A. Griffin and T. Nikuni, Phys. Rev. A,
in press; cond-mat/9705134. 

\bibitem{and97} M.R. Andrews {\it et al.}, Phys. Rev. Lett. {\bf 79},
553 (1997).

\bibitem{kps}A useful discussion of the conditions which must be
satisfied to probe the hydrodynamic region is given by
M. Kavoulakis, C.J. Pethick and H. Smith,
cond-mat/9710130.

\bibitem{uu33}E.A. Uehling and G.E. Uhlenbeck,  
Phys. Rev. {\bf 43}, 552 (1933).

\bibitem{ueh34}E.A. Uehling,  
Phys. Rev. {\bf 46}, 917 (1934).

\bibitem{kps2}
M. Kavoulakis, C.J. Pethick and H. Smith,
private communication.

\bibitem{kb}L.P. Kadanoff and G. Baym, {\it Quantum
Statistical Mechanics} (Benjamin, N.Y., 1962), Ch. 6.

\bibitem{kd}T.R. Kirkpatrick and J.R. Dorfman, J. Low Temp.
Phys. {\bf 58}, 304 (1985); {\it ibid}. 399 (1985).

\bibitem{fk}For a detailed review of the Chapman-Enskog method as
applied to classical gases, see J.H. Ferziger and H.G. Kaper,
{\it Mathematical Theory of Transport Processes in Gases}
(North-Holland, London, 1972).

\bibitem{huang}K. Huang, {\it Statistical Mechanics} (Wiley, N.Y., 1987),
2nd ed., p.113.

\bibitem{ll}See L.D. Landau and E.M. Lifshitz, 
{\it Fluid Mechanics} (Pergamon, Oxford, 1959), p.298.

\bibitem{footnote}
At $T_{\rm BEC}$ (when $z=1$ and $n_c=0$), our integral equations (22) 
are equivalent to Eqs. (24) in the second paper in Ref.~\cite{kd}, 
and hence our results
for $\kappa$ and $\eta$ should coincide at $T_{\rm BEC}$ 
with Eqs.~(26) of this reference.
However, our numerical factors differ slightly.

\bibitem{gps}S. Giorgini, L. Pitaevskii and S. Stringari,
J Low Temp. Phys., in press,
cond-mat/9704014.

\bibitem{khala}I.M. Khalatnikov, {\it An Introduction to the Theory of
Superfluidity} (Benjamin, N.Y., 1965).

\bibitem{cl}For example, see P.M. Chaikin and T.C. Lubensky,
{\it Principles of Condensed Matter Physics}
(Cambridge University Press, N.Y., 1995),
Ch. 8 and especially p.460ff.

\bibitem{hs}See, for example, T.L. Ho and V.B. Shenoy, cond-mat/9710275.
 
\end{references}
\end{document}